\documentclass{article}
\usepackage[preprint]{spconfa4}
\usepackage{amsmath,amsfonts,graphicx}
\usepackage{empheq}
\usepackage{siunitx}
\usepackage{cite}
\usepackage[english]{babel}
\usepackage{dblfloatfix}
\usepackage{anyfontsize}
\usepackage[subtle,title=normal,sections=normal,margins=normal,bibliography=normal]{savetrees}
\usepackage{blindtext}
\usepackage{pgfplots}
\usepackage{tikz}
\pgfplotsset{compat=newest}
\pgfplotsset{plot coordinates/math parser=false} 
\newlength\figureheight
\newlength\figurewidth 
\newcounter{MYtempeqncnt}
\newcommand{\vectorY}{\mathbf{y}\left(k,l\right)}
\newcommand{\vectorYherm}{\mathbf{y}^{H}\left(k,l\right)}
\newcommand{\vectorXs}{\mathbf{x}_{j}\left(k,l\right)}
\newcommand{\vectorXsDP}{\mathbf{x}_{d}^{\rm  DP}\left(k,l\right)}
\newcommand{\vectorXsRev}{\mathbf{x}_{d}^{\rm  R}\left(k,l\right)}

\newcommand{\vectorXdom}{\mathbf{x}_{d}\left(k,l\right)}

\newcommand{\vectorN}{\mathbf{n}\left(k,l\right)}
\newcommand{\vectorU}{\mathbf{u}_{d}\left(k,l\right)}
\newcommand{\vectorUherm}{\mathbf{u}^{H}_{d}\left(k,l\right)}

\newcommand{\RTFvectorSpeechsTimeVaryingDOA}{\mathbf{g}_{d}\left(k,\theta_{d}\left(l\right)\right)}
\newcommand{\ATFvectorSpeechsTimeVaryingDOA}{\mathbf{a}_{d}\left(k,\theta_{d}\left(l\right)\right)}
\newcommand{\RTFvectorSpeechd}{\mathbf{g}_{d}\left(k,\theta_{d}\left(l\right)\right)}
\newcommand{\RTFvectorSpeechdherm}{\mathbf{g}^{H}_{d}\left(k,\theta_{d}\left(l\right)\right)}
\newcommand{\RTFvectorHat}{\hat{\mathbf{g}}\left(k,l\right)}
\newcommand{\RTFvectorHatKL}{\hat{\mathbf{g}}\left(k,l\right)}
\newcommand{\RTFvectorBAR}{\bar{\mathbf{g}}\left(k,\theta_{i}\right)}
\newcommand{\RTFvectorBARherm}{\bar{\mathbf{g}}^{H}\left(k,\theta_{i}\right)}
\newcommand{\eOne}{\mathbf{e}_{1}}
\newcommand{\eOneT}{\mathbf{e}^{T}_{1}}
\newcommand{\phiY}{\boldsymbol{\Phi}_{\rm y}\left(k,l\right)}
\newcommand{\phiXd}{\boldsymbol{\Phi}_{\mathrm{x}_{d}}^{\rm{DP}}\left(k,\theta_{d}\left(l\right)\right)}

\newcommand{\phiUd}{\boldsymbol{\Phi}_{\mathrm{u}}\left(k,l\right)}

\newcommand{\psdSpeechd}{\Phi_{X_{d}}^{\rm{DP}}\left(k,l\right)}
\newcommand{\phiYHat}{\hat{\boldsymbol{\Phi}}_{\rm y}\left(k,l\right)}
\newcommand{\phiYHatLnegOne}{\hat{\boldsymbol{\Phi}}_{\rm y}\left(k,l-1\right)}
\newcommand{\GammaYHatKL}{\widehat{\boldsymbol{\Gamma}}_{\rm y}\left(k,l\right)}
\newcommand{\GammaUTildeKL}{\widetilde{\boldsymbol{\Gamma}}_{\rm u}\left(k\right)}
\newcommand{\phiYHatWhite}{\hat{\boldsymbol{\Phi}}^{\rm w}_{\rm y}\left(k,l\right)}
\newcommand{\phiUHat}{\hat{\boldsymbol{\Phi}}_{\mathrm{u}}\left(k,l\right)}
\newcommand{\phiUHatLnegOne}{\hat{\boldsymbol{\Phi}}_{\mathrm{u}}\left(k,l-1\right)}
\newcommand{\phiUHatsqrt}{\hat{\boldsymbol{L}}_{\mathrm{u}}\left(k,l\right)}
\newcommand{\phiUHatsqrtHerm}{\hat{\boldsymbol{L}}^{H}_{\mathrm{u}}\left(k,l\right)}
\newcommand{\phiUHatsqrtInv}{\hat{\boldsymbol{L}}^{-1}_{\mathrm{u}}\left(k,l\right)}
\newcommand{\phiUHatsqrtInvHerm}{\hat{\boldsymbol{L}}^{-H}_{\mathrm{u}}\left(k,l\right)}
\DeclareMathOperator{\sinc}{sinc}

\copyrightnotice{978-1-6654-6867-1/22/\$31.00~\copyright2022 IEEE}
                   
\title{COHERENCE-BASED FREQUENCY SUBSET SELECTION FOR BINAURAL RTF-VECTOR-BASED DIRECTION OF ARRIVAL ESTIMATION FOR MULTIPLE SPEAKERS}
\name{Daniel Fejgin and Simon Doclo\thanks{This work was funded by the Deutsche Forschungsgemeinschaft (DFG, German Research Foundation) under Germany's Excellence Strategy - EXC 2177/1 - Project ID 390895286 and Project ID 352015383 - SFB 1330 B 2.}}
\address{University of Oldenburg, Department of Medical Physics and Acoustics\\ and Cluster of Excellence Hearing4all, Oldenburg, Germany}
\usepackage[nopdftrailerid=true]{pdfprivacy}
\begin{document}
\ninept
\maketitle
\begin{abstract}
Recently, a method has been proposed to estimate the direction of \mbox{arrival} (DOA) of a single speaker by minimizing the frequency-averaged \mbox{Hermitian} angle between an estimated relative transfer function (RTF) vector and a database of proto\-type anechoic RTF vectors. In this paper, we extend this method to multi-speaker localization by intro\-ducing the frequency-averaged Hermitian angle spectrum and selecting peaks of this spatial spectrum. To construct the Hermitian angle spectrum, we consider only a subset of frequencies, where it is likely that one speaker is dominant. We compare the effectiveness of the generalized magnitude squared \mbox{coherence} and two coherent-to-diffuse ratio (CDR) estimators as frequency selection criteria. Simulation results for esti\-mating the DOAs of two speakers in a reverberant environment with diffuse-like babble noise using binaural hearing devices show that using the \mbox{binaural} effective-coherence-based CDR estimate as a frequency selection criterion yields the best performance.
\end{abstract}
\section{INTRODUCTION}
\label{sec:intro}
In many speech communication applications, such as teleconferencing systems and hearing devices, estimating the direction of arrival (DOA) of speech sources in the acoustic scene is of crucial importance \cite{Huang2008}. In this paper we specifically consider binaural hearing devices, for which several learning- and non-learning-based methods for multi-speaker DOA estimation have been proposed, e.g., based on interaural time and level differences \cite{May2011}, generalized cross correlation functions \cite{Kayser2014}, or using the subspace-based multiple signal classification (MUSIC) approach \cite{Schmidt19886}. In this paper we consider relative transfer function (RTF) vectors, which have been used for single-speaker binaural DOA estimation \cite{Braun2015,Li2015,Yang2021,Fejgin2021}, or for multi-speaker DOA estimation \cite{Hammer2021} (although not specifically in the context of binaural hearing devices).

In \cite{Fejgin2021} we proposed an RTF-vector-based binaural DOA \mbox{estimation} method for a single speaker by selecting the direction for which the frequency-averaged Hermitian angle between the estimated RTF \mbox{vector} and a database of prototype anechoic RTF vectors is minimized. In this paper we extend the DOA estimation method from \cite{Fejgin2021} to the multi-speaker case. Assuming that the number of speakers $J$ is known, multi-speaker DOA estimation could in principle be simply achieved by selecting $J$ peaks of the frequency-averaged Hermitian angle spectrum. Instead of averaging the Hermitian angle over all frequency bins (as in \cite{Fejgin2021}), we consider only a subset of frequency bins, where it is likely that one speaker dominates over all other speakers, noise, and rever\-beration. Common criteria to perform frequency bin subset selection in the context of DOA estimation are based on, e.g., signal-to-noise ratio (SNR) \cite{Li2015,Tho2014}, onsets \cite{Tho2014}, and coherence-based quantities such as the coherent-to-diffuse ratio (CDR) \cite{Brendel2018,Evers2018,Lee2020}.

In this paper we compare the effectiveness of coherence-based quantities for frequency bin subset selection, more in particular, the ge\-neralized magnitude squared coherence from \cite{Ramirez2008} and two recently proposed \mbox{binaural} CDR estimators from \cite{Loellmann2020} which are based on quantities to which we refer to as binaural generalized coherence and effective coherence. This means that the Hermitian angle is only \mbox{averaged} using frequency bins where the estimated binaural coherence-based quantity exceeds a certain threshold. For an acoustic scenario with two static \mbox{speakers} in a rever\-berant room with diffuse-like babble noise we \mbox{analyze} the performance of the binaural DOA esti\-mation method using the proposed frequency-averaged Hermitian angle spectrum based on simu\-lations with measured binaural room impulse responses (BRIRs). We compare the effectiveness and the influence of the threshold of the coherence-based selection criteria for frequency bin subset selection. Experimental results show that using the binaural effective-coherence-based CDR estimate yields the best DOA estimation performance of all considered cohere-based quantities, both for the Hermitian angle spectrum as well as for the MUSIC spectrum.

\section{SIGNAL MODEL AND NOTATION}
\label{sec:signalModel}
We consider a binaural hearing aid setup with $M$ microphones, i.e., $M/2$ microphones on each hearing aid. We consider an acoustic scenario with $J$ simultaneously active speakers $S_{1:J}$ located at DOAs $\theta_{1:J}$ (in the azimuthal plane) in a noisy and reverberant environment, where $J$ is assumed to be known. In the STFT domain, the $m$-th microphone signal can be written as
\begin{equation}
	Y_{m}\left(k,l\right) = \sum_{j=1}^{J}X_{m,j}\left(k,l\right) + N_{m}\left(k,l\right)\,, \quad m \in \left\{1,\dots,M\right\}
	\label{eq:signalModel_micComponent}
\end{equation}
where $k\in\left\{1,\dotsc,K\right\}$ and $l\in\left\{1,\dotsc,L\right\}$ denote the frequency bin index and the frame index, respectively, and $X_{m,j}\left(k,l\right)$ and $N_{m}\left(k,l\right)$ denote the $j$-th speech component and the noise component in the $m$-th microphone, respectively. Assuming one dominant speaker per time-frequency bin (indexed by $d$) and stacking all microphone signals in an $M$-dimensional vector $\vectorY =\left[Y_{1}\left(k,l\right),\,\dots,\,Y_{M}\left(k,l\right)\right]^{T}$, where $\left(\cdot\right)^{T}$ denotes transposition, the vector $\vectorY$ is given by
\begin{equation}
	\vectorY = \sum_{j=1}^{J}\vectorXs + \vectorN \approx \vectorXdom + \vectorN\,,
	\label{eq:signalModel_vectorized}
\end{equation} 
with $\vectorXs$, $\vectorXdom$, and $\vectorN$ defined similarly as $\vectorY$.

Assuming that the speech component for each (dominant) speaker $\vectorXdom$ can be split into a direct-path component $\vectorXsDP$ and a reverberant component $\vectorXsRev$ and assuming that the multiplicative transfer function approximation \cite{Avargel2007} holds for the direct-path component, $\vectorXdom$ can be written as
\begin{equation}
	\vectorXdom = \vectorXsDP +  \vectorXsRev = \ATFvectorSpeechsTimeVaryingDOA S_{d}\left(k,l\right) + \vectorXsRev\,
	\label{eq:domSpeech_ATF}
\end{equation}
where $\ATFvectorSpeechsTimeVaryingDOA$ denotes the direct-path acoustic transfer function (ATF) vector between the dominant speaker with DOA $\theta_{d}\left(l\right)$ and the micro\-phones. Choosing the first microphone as the reference microphone (without loss of generality), $\vectorXdom$ can also be written as
\begin{equation}
	\vectorXdom = \RTFvectorSpeechsTimeVaryingDOA X_{1,d}^{\rm  DP}\left(k,l\right) + \vectorXsRev\,,
	\label{eq:domSpeech_RTF}
\end{equation}
where 
\begin{equation}
	\RTFvectorSpeechsTimeVaryingDOA = \left[1, G_{2,d}\left(k,\theta_{d}\left(l\right)\right),\dots, G_{M,d}\left(k,\theta_{d}\left(l\right)\right)\right]^{T}
	\label{eq:definitionRTFvec}
\end{equation}
denotes the direct-path RTF vector and $X_{1,d}^{\rm  DP}\left(k,l\right)$ denotes the direct-path speech component of the dominant speaker in the reference microphone. The noise and reverberation components are condensed into the undesired component $\vectorU = \vectorN + \vectorXsRev$ such that $\vectorY \approx \vectorXsDP + \vectorU$.

Assuming uncorrelated direct-path speech and undesired components, the covariance matrix of the noisy microphone signals can be written as
\begin{equation}
	\phiY = \mathcal{E}\left\{\vectorY \vectorYherm\right\} = \phiXd + \phiUd\,,
	\label{eq:definitionPhiy}
\end{equation}
with 
\begin{align}
	\phiXd &= \RTFvectorSpeechd \RTFvectorSpeechdherm \psdSpeechd \label{eq:definitionPhiX}\,,\\
	\phiUd &= \mathcal{E}\left\{\vectorU \vectorUherm\right\}\label{eq:definitionPhiU}\,,
\end{align}
where $\left(\cdot\right)^{H}$ and $\mathcal{E}\left\{\cdot\right\}$ denote the complex trans\-position and expec\-tation operators, respectively. $\phiXd$ and $\phiUd$ denote the covariance matrices of the direct-path dominant speech component and undesired component, respectively, and $\psdSpeechd =\mathcal{E}\left\{\lvert X_{1,d}^{\rm  DP}\left(k,l\right)\rvert^{2}\right\}$ denotes the power spectral density of the direct-path dominant speech component in the \mbox{reference} microphone. 

\begin{figure*}[b]
	\hrule
	\vskip2.5pt
	\normalsize
	\setcounter{MYtempeqncnt}{\value{equation}}
	\setcounter{equation}{26}
	\begin{equation}
		\label{eq:SchwarzCDR}
		%
		f\left(\widehat{\Gamma}_{\rm y,eff},\widetilde{\Gamma}_{\mathrm{u}_{i^{\prime},j^{\prime}}}\right) = \frac{\widetilde{\Gamma}_{\mathrm{u}_{i^{\prime},j^{\prime}}}\, \Re\{\widehat{\Gamma}_{\rm y,eff}\} -\lvert\widehat{\Gamma}_{\rm y,eff}\rvert^{2} - \sqrt{{\widetilde{\Gamma}_{\mathrm{u}_{i^{\prime},j^{\prime}}}}^2\, {\Re\{\widehat{\Gamma}_{\rm y,eff}\}}^{2} - {\widetilde{\Gamma}_{\mathrm{u}_{i^{\prime},j^{\prime}}}}^{2}\, {\lvert\widehat{\Gamma}_{\rm y,eff}\rvert}^{2} + {\widetilde{\Gamma}_{\mathrm{u}_{i^{\prime},j^{\prime}}}}^{2} - 2\, \widetilde{\Gamma}_{\mathrm{u}_{i^{\prime},j^{\prime}}}\, \Re\{\widehat{\Gamma}_{\rm y,eff}\} + {\lvert\widehat{\Gamma}_{\rm y,eff}\rvert}^2}}{{\lvert\widehat{\Gamma}_{\rm y,eff}\rvert}^{2} - 1}
	\end{equation}
	\setcounter{equation}{\value{MYtempeqncnt}}
\end{figure*} 

\section{BINAURAL RTF-VECTOR-BASED DOA ESTIMATION}
\label{sec:doaEstimation}
To estimate the DOAs $\theta_{1:J}$ of all speakers, in this section we propose a multi-speaker extension of the binaural RTF-vector-based single-speaker DOA estimation method from \cite{Fejgin2021}. In Section \ref{ssec:doaEstimation_1spk} we briefly explain the single-speaker DOA estimation method, where the DOA is estimated by comparing the estimated direct-path RTF vector with a database of prototype anechoic RTF vectors based on the Hermitian angle. In Section \ref{ssec:doaEstimation_2spk} we propose a method to estimate the DOAs of multiple speakers based on the frequency-averaged Hermitian angle spectrum, where we consider several binaural coherence-based quantities for frequency bin subset selection.

\subsection{Single-speaker DOA estimation}
\label{ssec:doaEstimation_1spk}
To obtain an estimate of the direct-path RTF vector of the dominant speaker in each time-frequency bin, we use the state-of-the-art covariance whitening (CW) method \cite{Markovich2009}. First, the estimated noisy covariance matrix $\phiYHat$ is prewhitened using a square-root decomposition (e.g., Cholesky decomposition) of the estimated covariance matrix $\phiUHat$ of the undesired component , i.e.,
\begin{align}
	\phiUHat &= \phiUHatsqrt\phiUHatsqrtHerm\,,\\
	\phiYHatWhite &= \phiUHatsqrtInv\phiYHat\phiUHatsqrtInvHerm\,.
\end{align}
The direct-path RTF vector is then estimated as the normalized de\-whitened principal eigenvector of the pre\-whitened noisy covariance matrix, i.e.
\begin{equation}
	\label{eq:CW}
	\RTFvectorHat = \frac{\phiUHatsqrt\mathcal{P}\left\{\phiYHatWhite\right\}}{\eOneT\phiUHatsqrt\mathcal{P}\left\{\phiYHatWhite\right\}}\,,
\end{equation}
where $\mathcal{P}\left\{\cdot\right\}$ denotes the principal eigenvector of a matrix and $\eOne = \left[1,0,\dots,0\right]^{T}$ is an $M$-dimensional selection vector.

For each time-frequency bin, the estimated direct-path RTF vector $\RTFvectorHatKL$ is compared against a database of prototype an\-echoic RTF vectors $\RTFvectorBAR$ for different discrete directions $\theta_{i}\,, i=1,...,I$ using the so-called Hermitian angle \cite{Scharnhorst2001}, i.e.,
\begin{equation}
	\label{eq:hermitAngle_narrowband}
	p\left(k,l,\theta_{i}\right) = \arccos\left(\frac{\lvert\RTFvectorBARherm\RTFvectorHatKL\rvert}
	{\lVert\RTFvectorBAR\rVert_{2}\,\lVert\RTFvectorHatKL\rVert_{2}}\right)\,.
\end{equation} 
The DOA of the speaker is then estimated as the direction for which the Hermitian angle averaged over all frequencies (except DC) is maximized, i.e.,
\begin{equation}
	\label{eq:doaEstimate1spk}
	P^{\prime}\left(l,\theta_{i}\right)=-\sum_{k=2}^{K}~p\left(k,l,\theta_{i}\right)\,, \quad \hat{\theta}_{1}\left(l\right) =  \mathrm{argmax}_{\theta_{i}}\,P^{\prime}\left(l,\theta_{i}\right)\,.
\end{equation}

\subsection{Multi-speaker DOA estimation}
\label{ssec:doaEstimation_2spk}
When $J$ speakers are simultaneously active, the DOAs $\theta_{1:J}$ could in \mbox{principle} be estimated by selecting $J$ peaks of the frequency-averaged Hermitian angle $P^{\prime}\left(l,\theta_{i}\right)$ in \eqref{eq:doaEstimate1spk}. However, it should be realized that not all time-frequency bins are dominated by one speaker. Aiming at \mbox{including} only time-frequency bins where the estimated RTF vector in \eqref{eq:CW} is a good estimate for the direct-path RTF vector in \eqref{eq:definitionRTFvec} (of one of the speakers), we consider only a subset $\mathcal{K}\left(l\right)$ of frequency bins, for which it is likely that the direct-path of one speaker dominates over all other speakers, noise and rever\-beration. We define the frequency-averaged Hermitian angle spectrum as
\begin{equation}
	\boxed{P\left(l,\theta_{i}\right)=-\sum_{k\in\mathcal{K}\left(l\right)}p\left(k,l,\theta_{i}\right)}\label{eq:defHermitianAngleSpectrum}
\end{equation}
The DOAs $\hat{\theta}_{1:J}\left(l\right)$ are estimated by determining the $J$ peaks of this spatial spectrum (assuming $J$ be known).

To determine the subset $\mathcal{K}\left(l\right)$, several selection criteria have been proposed \cite{Li2015,Tho2014,Brendel2018,Evers2018,Lee2020}, many of which are coherence-based. More in particular, in this paper we consider the generalized magnitude squared coherence (GMSC) \cite{Ramirez2008} as well as two binaural CDR estimates presented in \cite{Loellmann2020} to which we refer to as binaural generalized-coherence-based CDR estimate and binaural effective-coherence-based CDR estimate.

According to \cite{Ramirez2008}, the generalized coherence generalizes the notion of coherence to $M\geq 2$ microphone signals and is defined as
\begin{equation}
	\widehat{\mathrm{GC}}\left(k,l\right) = \frac{\lambda_{\rm max}\left\{\GammaYHatKL\right\} -1}{M-1}\,,\label{eq:defGC}
\end{equation}
with $\GammaYHatKL$ containing the estimated coherence between the microphone signals, i.e.,
\begin{equation}
	\widehat{\Gamma}_{\mathrm{y}_{i,j}}\left(k,l\right) = \hat{\Phi}_{\mathrm{y}_{i,j}}\left(k,l\right)/\sqrt{\hat{\Phi}_{\mathrm{y}_{i,i}}\left(k,l\right)~\hat{\Phi}_{\mathrm{y}_{j,j}}\left(k,l\right)}
\end{equation} 
and $\lambda_{\rm max}\left\{\cdot\right\}$ denoting the principal eigenvalue of a matrix. The generalized magnitude squared coherence (GMSC) is then obtained as 
\begin{equation}
	\widehat{\mathrm{GMSC}}\left(k,l\right) = \left\lvert\widehat{\mathrm{GC}}\left(k,l\right)\right\rvert^{2}\,.
	\label{eq:defGMSC}
\end{equation}

In \cite{Loellmann2020} two binaural CDR estimates have been proposed. The first estimate, to which we refer to as binaural generalized-coherence-based CDR estimate, is defined as
\begin{equation}
	\widehat{\mathrm{CDR}}_{1}\left(k,l\right) = \frac{\widetilde{\mathrm{GC}}_{\rm u}\left(k\right) - \widehat{\mathrm{GC}}\left(k,l\right)}{\widehat{\mathrm{GC}}\left(k,l\right) - 1}\,,	\label{eq:defCDR1}
\end{equation}
where the (time-invariant) generalized coherence of the undesired \mbox{component} $\widetilde{\mathrm{GC}}_{\rm u}\left(k\right)$ is obtained similarly as $\widehat{\mathrm{GC}}\left(k,l\right)$ in \eqref{eq:defGC}, but using $\GammaUTildeKL$ instead of $\GammaYHatKL$. The model coherence matrix $\GammaUTildeKL$ models the coherence of the undesired component $\vectorU$ as a diffuse sound field, assuming that both the noise component $\vectorN$ as well as the reverberation component $\vectorXsRev$ can be modeled as a diffuse sound field. Depending on the considered microphone pair, either a free-field sinc-model \cite{Cook1955} or a modified sinc-model accounting for head shadow effects \cite{Lindevald1986} is employed 
\begin{align}
	\widetilde{\Gamma}_{\mathrm{u}_{i,j}}\left(k\right) &= \sinc{\left(\frac{\omega_{k}d_{i,j}}{c}\right)}\,,\\
	\widetilde{\Gamma}_{\mathrm{u}_{i^{\prime},j^{\prime}}}\left(k\right) &= \sinc{\left(\alpha\frac{\omega_{k}d_{i^{\prime},j^{\prime}}}{c}\right)}\,\frac{1}{\sqrt{1 + \left(\beta\frac{\omega_{k}d_{i^{\prime},j^{\prime}}}{c}\right)^{4}}}\label{eq:modifiedSincCoherence}\,,
\end{align}
where $\omega_{k}$ denotes the discrete angular frequency, $d_{i,j}$ denotes the distance between microphones $i$ and $j$, $c$ denotes the speed of sound, and $\alpha=0.5$ and $\beta=2.2$ \cite{Lindevald1986}.

The second estimate, to which we refer to as binaural effective-coherence-based CDR estimate, is defined as:
\begin{align}
	\widehat{\mathrm{CDR}}_{2}\left(k,l\right)&=f\left(\widehat{\Gamma}_{\rm y,eff}\left(k,l\right),~\widetilde{\Gamma}_{\mathrm{u}_{i^{\prime},j^{\prime}}}\left(k\right)\right)\,,	\label{eq:defCDR2}\\
	\widehat{\Gamma}_{\rm y, eff}\left(k,l\right) &= \frac{1}{\vert\mathcal{M}\vert}\sum_{i,j\in\mathcal{M}}\widehat{\Gamma}_{\mathrm{y}_{i,j}}\left(k,l\right)\,,\label{eq:effectiveCoherence}
\end{align}
where the effective coherence $\hat{\Gamma}_{\rm y, eff}\left(k,l\right)$ in \eqref{eq:effectiveCoherence} represents the average coherence between all possible microphone pairs between the left and the right hearing aid (denoted as the microphone set $\mathcal{M}$), and the CDR functional $f$ in \eqref{eq:SchwarzCDR} has been introduced in \cite{Schwarz2015}.

Given the coherence-based quantities $\widehat{\mathrm{GMSC}}\left(k,l\right)$, $\widehat{\mathrm{CDR}}_{1}\left(k,l\right)$, and $\widehat{\mathrm{CDR}}_{2}\left(k,l\right)$, we propose to define the subset $\mathcal{K}\left(l\right)$ of frequency bins to be included in the computation of the Hermitian angle spectrum in \eqref{eq:defHermitianAngleSpectrum} as
\begin{empheq}[box=\fbox]{align}
	\mathcal{K}_{\rm GMSC}\left(l\right) &= \left\{k: \widehat{\mathrm{GMSC}}\left(k,l\right)\geq\mathrm{GMSC}_{\rm min}\right\}\label{eq:freqSelectionRule1}\\
	\mathcal{K}_{\mathrm{CDR}_{1,2}}\left(l\right) &= \left\{k: \widehat{\mathrm{CDR}}_{1,2}\left(k,l\right)\geq\mathrm{CDR}_{\rm min}\right\}\label{eq:freqSelectionRule2}
\end{empheq}
where $\mathrm{GMSC}_{\rm min}$ and $\mathrm{CDR}_{\rm min}$ denote frequency- and frame-in\-de\-pendent thresholds.

\subsection{Baseline method: MUSIC}
As baseline method for multi-speaker DOA estimation we consider MUSIC \cite{Schmidt19886}, which is based on the orthogonality between the acoustic transfer function vector $\ATFvectorSpeechsTimeVaryingDOA$ and the noise subspace of $\phiY$ (see \eqref{eq:domSpeech_ATF} and \eqref{eq:definitionPhiy}). The narrowband MUSIC cost function is given by
\begin{equation}
	\label{eq:MUSIC_narrowband}
	p^{\left(\rm MUSIC\right)}\left(k,l,\theta_{i}\right) = \frac{1}{\Vert\bar{\mathbf{a}}^{H}\left(k,\theta_{i}\right)\hat{\boldsymbol{Q}}_{\mathrm{y,u}}\left(k,l\right)\Vert_{2}^{2}}\,.
\end{equation}
with $\bar{\mathbf{a}}\left(k,\theta_{i}\right)$ denoting the prototype anechoic acoustic transfer function vector for direction $\theta_{i}$ and $\hat{\boldsymbol{Q}}_{\mathrm{y,u}}\left(k,l\right)$ denoting the estimated noise subspace of $\phiYHat$. Performing the incoherent frequency averaging method as described in \cite{Salvati2014} and considering frequency bin subset selection as in Section \ref{ssec:doaEstimation_2spk}, the frequency-averaged normalized MUSIC spectrum is defined as
\begin{equation}
	P^{\left(\rm MUSIC\right)}\left(l,\theta_{i}\right)=\sum_{k\in\mathcal{K}\left(l\right)}\frac{p^{\left(\rm MUSIC\right)}\left(k,l,\theta_{i}\right)}{\underset{\theta_{j}}{\rm max}~p^{\left(\rm MUSIC\right)}\left(k,l,\theta_{j}\right)}\label{eq:defBroadbandMUSIC}\,.
\end{equation}
The DOAs $\hat{\theta}^{\left(\rm MUSIC\right)}_{1:J}\left(l\right)$ are estimated by determining the $J$ peaks of this spatial spectrum (assuming $J$ to be known).

\section{EXPERIMENTAL RESULTS}
\label{sec:experiments}
For an acoustic scenario with two static speakers in a rever\-berant room with diffuse-like babble noise, in this section we compare the \mbox{performance} of the coherence-based selection criteria discussed in \mbox{Section} \ref{ssec:doaEstimation_2spk}, both for the Hermitian angle spectrum as well as for the baseline MUSIC spectrum. The experimental setup and implementation details of the algorithms are described in Section \ref{ssec:experiments_details}. The results in terms of localization accuracy are presented and discussed in Section \ref{ssec:experiments_results}.

\subsection{Experimental setup and implementation details}
\label{ssec:experiments_details}
To simulate the binaural microphone signals, we use measured binaural room impulse responses (BRIRs) from the \verb|office_i| scenario (rever\-beration time $T_{60} \approx \SI{300}{\milli\second}$) of the database of \cite{Kayser2009}. This database contains measured BRIRs for DOAs in the range $\left[\SI{-80}{\degree}\,,\SI{80}{\degree}\right]$ with an angular resolution of \SI{5}{\degree}. Although the used hearing aids contain three micro\-phone each, we only consider the front and rear microphones on each hearing aid ($M=4$). We simulate several static two-speaker sce\-narios ($J=2$), where for each possible DOA combination with a minimum angular spacing of \SI{15}{\degree} (in total 930 DOA combinations) clean speech signals (male, female) from the DNS Challenge dataset \cite{Reddy2020} are convolved with the corresponding BRIRs. The speech signals are constantly active and are approximately $\SI{4}{\second}$ long. The average broadband speech power across all microphones is set to the same value for both speakers. Diffuse-like multi-channel babble noise is generated using the method in \cite{Habets2008} and added to the speech components of the microphone signals. The signal-to-noise ratio (SNR) is set to $\left\{0\,,5\,,20\right\}$\,\si{\decibel}, where the SNR is defined as the average broadband speech power of one speaker across all microphones to the average broadband noise power across all microphones. The microphone signals are simulated at a sampling rate of $\SI{16}{\kilo\hertz}$.

The simulated microphone signals are processed in the STFT domain using \SI{32}{\milli\second} square-root Hann windows with \SI{50}{\percent} overlap. The anechoic BRIRs from \cite{Kayser2009} with an angular resolution of \SI{5}{\degree} in the range $\left[\SI{-180}{\degree}\,,\SI{175}{\degree}\right]$ are used to generate the database of prototype anechoic ATF vectors $\left\{\bar{\mathbf{a}}\left(k,\theta_{i}\right)\right\}_{i=1}^{I}$ and RTF vectors $\left\{\RTFvectorBAR\right\}_{i=1}^{I}$, with $I=72$. For each time-frequency bin the noisy covariance matrix $\phiY$ and the undesired covariance matrix $\phiUd$ are estimated using recursive smoothing during speech-and-noise periods and noise-only periods, respectively as
\begin{align}
	\addtocounter{equation}{1}
	\phiYHat &= \alpha_{\rm y}\phiYHatLnegOne + \left(1-\alpha_{\rm y}\right) \mathbf{y}\left(k,l\right)\mathbf{y}^{H}\left(k,l\right)\label{eq:SOS_RyUpdate}\\
	\phiUHat &= \alpha_{\rm u}\phiUHatLnegOne + \left(1-\alpha_{\rm u}\right) \mathbf{y}\left(k,l\right)\mathbf{y}^{H}\left(k,l\right)\label{eq:SOS_RnUpdate}\,,
\end{align}
where the smoothing factors $\alpha_{\rm y}$ and $\alpha_{\rm u}$ correspond to time constants of \SI{250}{\milli\second} and \SI{500}{\milli\second}, respectively. The speech-and-noise periods and noise-only periods are determined based on the thresholded speech presence probability \cite{Gerkmann2012}, averaged over all microphones.

Performance is assessed in terms of the localization accuracy, i.e. the percentage of correctly localized frames. Similarly as in \cite{Hammer2021}, we consider a frame to be correctly localized only if both estimated DOAs are within $\pm\SI{5}{\degree}$ of the true DOAs.

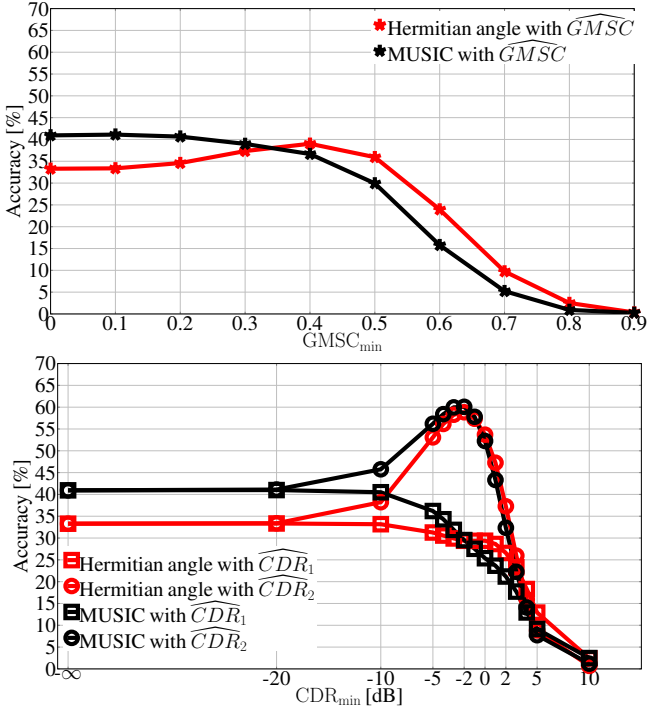
\begin{figure}[t]%
	\centering
	\begin{tikzpicture}[scale=0.195]

\begin{axis}[%
width=15.5in,
height=8.175in,
at={(2.6in,1.103in)},
scale only axis,
xmin=0,
xmax=0.9,
xtick={{0},{0.1},{0.2},{0.3},{0.4},{0.5},{0.6},{0.7},{0.8},{0.9}},
xticklabels={{0},{0.1},{0.2},{0.3},{0.4},{0.5},{0.6},{0.7},{0.8},{0.9}},
xticklabel style={font=\color{black}, font = \fontsize{40}{1}\selectfont},
xlabel style={font=\color{black}, font = \fontsize{40}{1}\selectfont},
xlabel={$\mathrm{GMSC}_{\rm min}$},
ymin=0,
ymax=70,
ytick={{0},{5},{10},{15},{20},{25},{30},{35},{40},{45},{50},{55},{60},{65},{70},{75},{80},{85},{90},{95},{100}},
yticklabels={{0},{5},{10},{15},{20},{25},{30},{35},{40},{45},{50},{55},{60},{65},{70},{75},{80},{85},{90},{95},{100}},
yticklabel style={font=\color{black}, font = \fontsize{40}{1}\selectfont},
ylabel style={font=\color{black}, font = \fontsize{40}{1}\selectfont},
ylabel={Accuracy [\%]},
axis background/.style={fill=white},
xmajorgrids,
ymajorgrids,
legend style={at={(0.55,0.99)}, anchor=north west, legend cell align=left, align=left, draw=none, 
	font = \fontsize{40}{1}\selectfont,
	row sep=3pt}
]
\addplot [color=red, line width=8.0pt, mark size=12pt, mark=star, mark options={solid, red}]
  table[row sep=crcr]{%
0	33.2896977862911\\
0.100000000000001	33.3782082549251\\
0.200000000000003	34.5884300299622\\
0.299999999999997	37.3403385070599\\
0.399999999999999	39.0169092780367\\
0.5	35.9430994594604\\
0.600000000000001	23.9463679557468\\
0.700000000000003	9.74887296353921\\
0.799999999999997	2.5116816785496\\
0.899999999999999	0.335290677442792\\
};
\addlegendentry{Hermitian angle with $\widehat{GMSC}$}

\addplot [color=black, line width=8.0pt, mark size=12pt, mark=star, mark options={solid, black}]
  table[row sep=crcr]{%
0	40.9342486905119\\
0.100000000000001	41.1123853838629\\
0.200000000000003	40.6570996613366\\
0.299999999999997	38.9817412063658\\
0.399999999999999	36.6665227349801\\
0.5	29.9322052495914\\
0.600000000000001	15.7781939407649\\
0.700000000000003	5.22833451562781\\
0.799999999999997	0.924615699698684\\
0.899999999999999	0.248721049011614\\
};
\addlegendentry{MUSIC with $\widehat{GMSC}$}

\end{axis}

\end{tikzpicture}%
%
	\vspace{0mm}
	\begin{tikzpicture}[scale=0.195]

\begin{axis}[%
width=15.5in,
height=8.175in,
at={(2.6in,1.103in)},
scale only axis,
xmin=-41,
xmax=15,
xtick={{-40},{-20},{-10},{-5},{-2},{0},{2},{5},{10}},
xticklabels={{-$\infty$},{-20},{-10},{-5},{-2},{0},{2},{5},{10}},
xticklabel style={font=\color{black}, font = \fontsize{40}{1}\selectfont},
xlabel style={font=\color{black}, font = \fontsize{40}{1}\selectfont},
xlabel={$\mathrm{CDR}_{\rm min}$ [dB]},
ymin=0,
ymax=70,
ytick={{0},{5},{10},{15},{20},{25},{30},{35},{40},{45},{50},{55},{60},{65},{70},{75},{80},{85},{90},{95},{100}},
yticklabels={{0},{5},{10},{15},{20},{25},{30},{35},{40},{45},{50},{55},{60},{65},{70},{75},{80},{85},{90},{95},{100}},
yticklabel style={font=\color{black}, font = \fontsize{40}{1}\selectfont},
ylabel style={font=\color{black}, font = \fontsize{40}{1}\selectfont},
ylabel={Accuracy [\%]},
axis background/.style={fill=white},
xmajorgrids,
ymajorgrids,
legend style={at={(0.01,0.4)}, anchor=north west, legend cell align=left, align=left, draw=none, font = \fontsize{40}{1}\selectfont,row sep=3pt}
]
\addplot [color=red, line width=8.0pt, mark size=12pt, mark=square, mark options={solid, red}]
  table[row sep=crcr]{%
-40	33.2896977862911\\
-20	33.3369035029065\\
-10	33.1677571893483\\
-5	31.2673979913376\\
-4	30.5681192520527\\
-3	29.9059202261684\\
-2	29.3693310140978\\
-1	29.3648044217758\\
0	29.3854794624746\\
1	28.6331644656339\\
2	26.5732421744232\\
3	23.3057263975748\\
4	18.3113984064873\\
5	12.9085350162597\\
10	2.41596175407273\\
};
\addlegendentry{Hermitian angle with $\widehat{CDR}_{1}$}

\addplot [color=red, line width=8.0pt, mark size=12pt, mark=o, mark options={solid, red}]
  table[row sep=crcr]{%
-40	33.2896977862911\\
-20	33.3830274730889\\
-10	38.212739973131\\
-5	53.0757275215111\\
-4	56.1510107399703\\
-3	58.2674156140726\\
-2	58.832437512908\\
-1	57.3471545328056\\
0	53.6971544867812\\
1	47.268086411799\\
2	37.2780127424708\\
3	25.9680408644707\\
4	15.9737634219718\\
5	8.82645842587539\\
10	0.688794992408212\\
};
\addlegendentry{Hermitian angle with $\widehat{CDR}_{2}$}

\addplot [color=black, line width=8.0pt, mark size=12pt, mark=square, mark options={solid, black}]
  table[row sep=crcr]{%
-40	40.9342486905119\\
-20	40.9951450394065\\
-10	40.5000101578842\\
-5	36.1912749302271\\
-4	34.3079816838052\\
-3	31.8706915537386\\
-2	29.5351049264625\\
-1	27.5305926049298\\
0	25.3691788140725\\
1	23.6572324623861\\
2	21.1777091011268\\
3	17.7247375093229\\
4	12.9405348408629\\
5	9.14556196026192\\
10	2.48556181994194\\
};
\addlegendentry{MUSIC with $\widehat{CDR}_{1}$}

\addplot [color=black, line width=8.0pt, mark size=12pt, mark=o, mark options={solid, black}]
  table[row sep=crcr]{%
-40	40.9342486905119\\
-20	41.1139521225351\\
-10	45.7571175550611\\
-5	56.2001959547389\\
-4	58.4003062624843\\
-3	59.9594487720165\\
-2	60.0863134693816\\
-1	57.7997891064024\\
0	52.2507603161227\\
1	43.3162419306594\\
2	32.3522407255604\\
3	22.2534336771753\\
4	14.0665202833226\\
5	7.7192834178481\\
10	1.20072916502198\\
};
\addlegendentry{MUSIC with $\widehat{CDR}_{2}$}

\end{axis}

\end{tikzpicture}%
%
	\vspace{-3mm}
	\caption{Average localization accuracy using the Hermitian angle spectrum (red) and MUSIC spectrum (black), with frequency bin subset selection based on generalized magnitude squared coherence (top) or coherent-to-diffuse ratio (bottom).}
	\vskip-3.4mm
	\label{fig:res}
\end{figure}

\subsection{Results}
\label{ssec:experiments_results}
For the two-speaker scenarios described in Section \ref{ssec:experiments_details}, we investigate the localization accuracy of the multi-speaker DOA estimation methods proposed in Section \ref{ssec:doaEstimation_2spk}. More in particular, we compare the influence of selecting frequency bins based on several coherence-based quantities (GMSC and CDR). Since it is unrealistic to assume that the thresholds in \eqref{eq:freqSelectionRule1} and \eqref{eq:freqSelectionRule2} can be chosen scenario-dependent, the localization accuracy is averaged over all considered DOA combinations and SNRs (see Section \ref{ssec:experiments_details}).

For the Hermitian angle spectrum in \eqref{eq:defHermitianAngleSpectrum} and the MUSIC spectrum in \eqref{eq:defBroadbandMUSIC}, Fig. 1 depicts the average localization accuracy for \mbox{different} \mbox{thresholds} of either the generalized magnitude squared coherence $\widehat{\mathrm{GMSC}}$ in \eqref{eq:freqSelectionRule1} or the coherent-to-diffuse ratios $\widehat{\mathrm{CDR}}_{1}$ and $\widehat{\mathrm{CDR}}_{2}$ in $\eqref{eq:freqSelectionRule2}$. For all selection criteria, a small threshold corresponds to selecting many frequency bins, whereas a large threshold corresponds to selecting few frequency bins. When the threshold is equal to zero ($-\infty$\,\si{\decibel}), all frequency bins are selected. For the selection criteria $\widehat{\mathrm{GMSC}}$ and $\widehat{\mathrm{CDR}}_{1}$, it can be observed that setting the threshold $\mathrm{GMSC}_{\rm,min}> 0$ and $\mathrm{CDR}_{\rm min}>0$ does not enable to significantly increase the localization accuracy compared to selecting all frequency bins, both for the Hermitian angle spectrum as well as for the MUSIC spectrum. In contrast, for the selection criterion $\widehat{\mathrm{CDR}}_{2}$ based on the binaural effective coherence a significant influence of the CDR threshold can be observed, which is in line with \cite{Brendel2018,Evers2018,Lee2020}. DOA estimation using either the Hermitian angle \mbox{spectrum} or the MUSIC spectrum works best when choosing a threshold of $\mathrm{CDR}_{\rm min}\approx\SI{-2}{\decibel}$. Using this threshold, the localization accuracy can be increased from about $\SI{35}{\percent}$ to $\SI{60}{\percent}$ for the Hermitian angle spectrum and from about $\SI{40}{\percent}$ to $\SI{60}{\percent}$ for the MUSIC spectrum. When considering the SNR-dependent localization accuracies (not depicted here) for the $\widehat{\mathrm{CDR}}_{2}$ selection criterion, there is always a distinct peak at $\mathrm{CDR}_{\rm min}\approx\SI{-2}{\decibel}$ implying an SNR-independent optimal threshold value. Although the proposed Hermitian angle spectrum as a functional for DOA estimation does not outperform the MUSIC spectrum, it represents a viable alter\-native, especially given the fact that low-complexity RTF vector estimation methods have been proposed assuming the availability of one or more external microphones \cite{Goessling2018,Goessling2019}.


\section{CONCLUSIONS}
\label{sec:pagestyle}
In this paper we proposed an extension of a recently proposed RTF-vector-based DOA estimation method for a single speaker to RTF-vector-based DOA estimation for multiple speakers by introducing the Hermitian angle spectrum. To construct this spatial spectrum, we consider only a subset of frequency bins, where it is likely that one speaker dominates over all other speakers, noise, and reverberation. In this paper we compared the effectiveness of the generalized magnitude squared coherence, the binaural generalized-coherence-based estimate of the CDR, and the binaural effective-coherence-based estimate of the CDR as criteria for frequency bin subset selection. Using measured BRIRs, we simulated acoustic scenarios with two static speakers in a reverberant room with diffuse-like babble noise. Simulation results for DOA estimation using binaural hearing devices show no signi\-ficant increase in the localization accuracy when using either the ge\-neralized magnitude squared coherence or the binaural generalized-coherence-based estimate of the CDR as selection criteria compared to selecting all frequency bins. In contrast, the localization accuracy can be significantly increased when using the binaural effective-coherence-based estimate of the CDR as a selection criterion. Using the optimal threshold value of about $\SI{-2}{\decibel}$, enables to increase the average localization accuracy for the proposed Hermitian angle spectrum as a functional for DOA estimation from about $\SI{35}{\percent}$ to $\SI{60}{\percent}$ compared to when selecting all frequency bins. Using the same optimal threshold value when using the \mbox{MUSIC} spectrum as a functional for DOA estimation, localization accuracy can be increased from about $\SI{40}{\percent}$ to $\SI{60}{\percent}$ compared to when selecting all frequency bins. For the binaural effective-coherence-based estimate of the CDR as a frequency bin subset selection criterion, experimental results also imply an optimal threshold value that is independent of the SNR.

\bibliographystyle{IEEEbib}
\bibliography{refs}
\end{document}